\documentclass[modern]{aastex61}

\usepackage{graphicx}% Including figure files
\usepackage{amsmath}% Advanced maths commands
\usepackage{amssymb}% Extra maths symbols
\usepackage{enumitem}

\newcommand\mybar{\kern1pt\rule[-\dp\strutbox]{.8pt}{\baselineskip}\kern1pt}

\setlist[itemize]{noitemsep, topsep=0pt, leftmargin=*}

\shorttitle{A Hot Subdwarf Pulsar}
\shortauthors{Loeb and Maoz}

\begin{document}

\title{A Hot Subdwarf Model for the 18.18 Minute Pulsar GLEAM-X}

\author{Abraham Loeb}
\affiliation{Astronomy Department, Harvard University, 60 Garden
  St., Cambridge, MA 02138, USA}
\author{Dan Maoz}
\affiliation{School of Physics \& Astronomy, Tel Aviv University,
  Tel Aviv 69978, Israel}

\begin{abstract}
We suggest that the recently discovered, enigmatic pulsar with a
period of 18.18 minutes, GLEAM-X J162759.5-523504.3, is most likely a
hot subdwarf (proto white dwarf). A magnetic dipole model explains the
observed period and period-derivative for a highly magnetized ($\sim
10^8$G), hot subdwarf of typical mass $\sim 0.5M_\odot$ and radius
$\sim 0.3R_\odot$, and an age of $\sim 3\times 10^4$~yr. The subdwarf
spin is close to its breakup speed and its spindown luminosity is near
its Eddington limit, likely as a result of accretion from a companion.
\end{abstract}

\section*{Introduction}
A recent analysis of archival low-frequency radio data collected by
the Murchison Widefield Array (MWA), has revealed an unusual pulsar in
the Milky-Way, GLEAM-X J162759.5-523504.3, with a period of
$P=1091.1690(\pm 0.0005)~{\rm s}$ ($\approx$18.18 minutes) and a
best-fit period derivative of ${\dot P}=6\times 10^{-10}{\rm
  s~s^{-1}}$ \citep{2022Natur.601..526H}. The data set a maximum on
the potential spin-down luminosity of a neutron star source, ${\dot
  E}<1.2\times 10^{28}~{\rm erg~s^{-1}}$, which is a few thousand
times lower than the brightest pulses inferred for GLEAM-X, $4\times
10^{31}~{\rm erg~s^{-1}}$. As noted by the discovery team, the radio
luminosities of pulsars usually constitute only a small fraction of
their spindown luminosity, and therefore a neutron star origin for
this pulsar is untenable. Below we show that a hot subdwarf (HSD)
pulsar, a stellar core evolving towards a white dwarf state but not
yet fully degenerate, satisfies the observational constraints.

\section*{Model}

A misaligned rotator model for a pulsar consisting of a dipole
magnetic field, $B$, and radius, $R$, yields a spindown luminosity,
\begin{equation}
{\dot E}=\left({4\pi^4\over 9c^3}\right)\left({B^2R^6\over
  P^4}\right)=10^{38}~{\rm erg~s^{-1}}\left({B\over
    10^8{\rm G}}\right)^{2}\left({R\over 0.3R_\odot}\right)^{6} ,
\label{Edot}
\end{equation}
where we have adopted an average value of $\sim 1/3$ for
$\sin^2\alpha$, with $\alpha$ the angle between the magnetic moment
and the rotation axis \citep{1985bhwd.book.....S}.  Indeed, AR Scorpii
is a recently discovered 2-min period pulsar in a 3.5-hr orbit with an
M-type star, in which the pulsar is powered by the spindown of a white
dwarf with $B\lesssim 10^8$ G
\citep{2016Natur.537..374M,2017NatAs...1E..29B,2021ApJ...908..195G}.
Emulating that system but with typical HSD mass $M=0.5(\pm
0.1)~M_\odot$ \citep{2009A&A...504L..13Z,2021FrASS...8...19L} and
radius $R=0.25(\pm 0.15)R_\odot$
\citep{2009ARA&A..47..211H,2019NatAs...3..553R}, suggests that the
spindown luminosity could easily power the radio emission from
GLEAM-X.  Interestingly, the value of ${\dot E}$ for $B\sim 10^8$ G is
close to the Eddington luminosity limit for a HSD mass of $M\sim
0.5M_\odot$, possibly maintaining a puffed-up HSD radius of $R\sim
0.3R_\odot$. The lack of a strong X-ray or UV counterpart for the
radio detection may indicate that the HSD is young and still obscured
in these spectral bands by the envelope of its red giant
progenitor.

The spindown time in the magnetic dipole model is given by, 
\begin{equation}
{P\over {\dot P}}=\left({9 c^3\over 2\pi^2 }\right)\left({I P^2\over
  B^2 R^6}\right)=2.5\times 10^{12}~{\rm s}\left({M\over 0.5
  M_\odot}\right)\left({B\over 10^8{\rm G}}\right)^{-2}\left({R\over
  0.3R_\odot}\right)^{-4} ,
\label{pdot}
\end{equation}
where $I= (2/5) MR^2$ is the moment of inertia for a star of mass $M$
and radius $R$. The measured value of $(P/{\dot P})\approx 2\times
10^{12}~{\rm s}$, agrees naturally with the parameters of a highly
magnetized HSD. It implies an age of $\sim (P/2{\dot P})\sim 3\times
10^4$yr for the HSD pulsar, which is consistent with a stellar remnant
that has yet to cool to a white dwarf \citep{2021arXiv211000598F}.
HSDs with rotation periods similar to that of GLEAM-X are known to
exist \citep{2019ApJ...878L..35K}.

\section*{Conclusions}

The breakup period of a HSD,
\begin{equation}
P_{\rm break}\sim 2\left({GM\over R^3}\right)^{-1/2}= 800~{\rm
  s}\left({M\over 0.5 M_\odot}\right)^{-1/2}\left({R\over
  0.3R_\odot}\right)^{3/2} ,
\label{br}
\end{equation}
is comparable to the observed period of GLEAM-X,
making the observed rotation period physically plausible for a HSD
that was spun up close to its maximum possible spin by accretion from
a companion. The sporadic appearance of the observed pulses from the source,
perhaps during only $\sim 2\%$ of the time, also suggests the presence of a
companion star, which endowed the HSD pulsar with its rapid spin and
high magnetic field. 
%If the companion is tidally-locked with an
%orbital period equal to the observed spin period, $P$, it must be a
%compact object like another HSD, a white dwarf, a neutron star or a
%black hole near contact with the HSD. In that case, the gravitational
%wave emission from this system could potentially be detected by LISA
%\citep{2020ApJ...889...49B}.

In AR Scorpii
\citep{2016Natur.537..374M,2017NatAs...1E..29B,2021ApJ...908..195G},
the companion M-type star is not in synchronous orbit with the white
dwarf spin. A similar configuration in GLEAM-X would produce a
few-seconds modulation in the pulse arrival time owing to the light
travel time (Roemer delay), which in principle may be detectable in
the existing data. The absence of such modulation would suggest either
a lack of a companion (since birth, or after evolution -
e.g. explosive destruction or unbinding of, or merger with, the
companion) or a compact companion in an orbit that has been tidally
synchronized with the HSD spin.

\bigskip
\bigskip
\section*{Acknowledgements}

This work was supported in part by Harvard's {\it Black Hole
  Initiative}, which is funded by grants from JFT and GBMF (for A.L.).
D.M. acknowledges support by European Research Council (ERC) FP7 Grant
No. 833031, and grants from the Israel Science Foundation (ISF) and
the German Israeli Science Foundation (GIF).

\bibliographystyle{aasjournal}
\bibliography{wd}
\label{lastpage}
\end{document}